\documentclass{jps-cp}
\usepackage{txfonts} 
\usepackage{braket}
\usepackage{color}
\usepackage{amsmath}
\usepackage{ulem}

\title{Magnetic properties of the $S=1$ Kitaev model
  with anisotropic interactions}

\author{Tetsuya \textsc{Minakawa}$^{1}$, Joji \textsc{Nasu}$^{2}$
  and Akihisa \textsc{Koga}$^{1}$}

\inst{$^{1}$ Department of Physics, Tokyo Institute of Technology, Meguro, Tokyo 152- 8551, Japan
  \\
  $^{2}$ Department of Physics, Yokohama National University, Hodogaya, Yokohama 240-8501, Japan}

\recdate{\today}

\abst{
  We investigate magnetic properties in the $S=1$ Kitaev model in the anisotropic limit.
  Performing the fourth-order perturbation expansion with respect to the $x$-bonds, $y$-bonds, and
  magnetic field, we derive the effective Hamiltonian, where
  the low-energy physics should be described by the free spins with
  an effective magnetic field.
  Making use of the exact diagonalization method for small clusters,
  we discuss ground-state properties in the system complementary.
}

\begin{document}
\maketitle

\section{Introduction}
Recently, the Kitaev model~\cite{Kitaev} and its related models have attracted
much interest since the possibility of
the direction-dependent Ising interactions has been
proposed in the realistic materials\cite{Jackeli}.
One of the important issues in this field is the ground-state property
of Kitaev models
whose the spin is generalized to an arbitrary $S$~\cite{Baskaran,S1Koga,S1Oitmaa,Kee,MixKitaev}.
In our previous paper~\cite{Minakawa},
we have focused on the spin-$S$ Kitaev model in the anisotropic limit
to clarify that the effective Hamiltonian depends on the magnitude of spins.
When the spin is a half integer, the system is described by the toric code Hamiltonian
and the topological nature of the ground state has been discussed in detail.
On the other hand, in an integer-spin case, the effective Hamiltonian is equivalent to
a free spin model under a uniform magnetic field, where quantum fluctuations
are quenched.
In contrast to the former case,
ground-state properties in the spin-integer model
are little understood.
In particular, it is desired to discuss the effect of the magnetic field in the system.

In this paper, we study the $S=1$ Kitaev model under the uniform magnetic field
as a simplest model with integer spins
and discuss ground-state properties by means of the perturbation theory and
exact diagonalization (ED).

\section{Model and Results}

We consider the $S=1$ Kitaev model under an external magnetic field, which is given as,
\begin{equation}
  \mathcal{H} =
  -J_x\sum_{\langle i,j \rangle_x}S_i^x S_j^x
  -J_y\sum_{\langle i,j \rangle_y}S_i^y S_j^y
  -J_z\sum_{\langle i,j \rangle_z}S_i^z S_j^z
  - {\bf h} \cdot \sum_i {\bf S}_i,\label{H}
\end{equation}
where $S_i^\alpha$ is the $\alpha(=x,y,z)$ component of an $S=1$ operator at the $i$th site.
$J_\alpha$ is the exchange constant on the $\alpha(=x,y,z)$ bonds connecting the nearest-neighbor sites $\langle i,j \rangle_\alpha$.
Here, we set $g\mu_B=1$ and consider the magnetic field $h$
given by ${\bf h}=h(\cos\varphi, \sin\varphi, 0)$.
\begin{figure}[htb]
\centering
\includegraphics{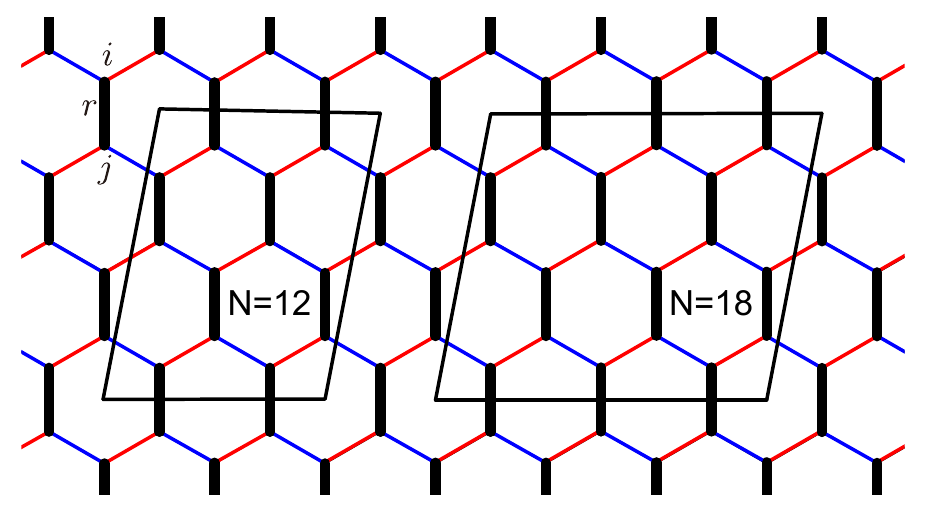}
\caption{
  Lattice structure of the Kitaev model.
  $x$ and $y$ bonds are expressed by the thin red and blue lines and $z$ bonds by the bold lines.
  Clusters used in the ED method are represented by
  the thin black lines.
}
\label{honeycomb}
\end{figure}
The model Hamiltonian is schematically shown in Fig.~\ref{honeycomb}.
It is known that, when no magnetic field is applied,
a local $Z_2$ symmetry in the generalized Kitaev model plays
an important role in realizing the quantum spin liquid state~\cite{Baskaran,MixKitaev}.
On the other hand, when $h\neq 0$, there never exists such a symmetry and thereby
it is unclear how stable the quantum spin liquid is against
the magnetic field.

\begin{figure}[htb]
\centering
\includegraphics[width=10cm]{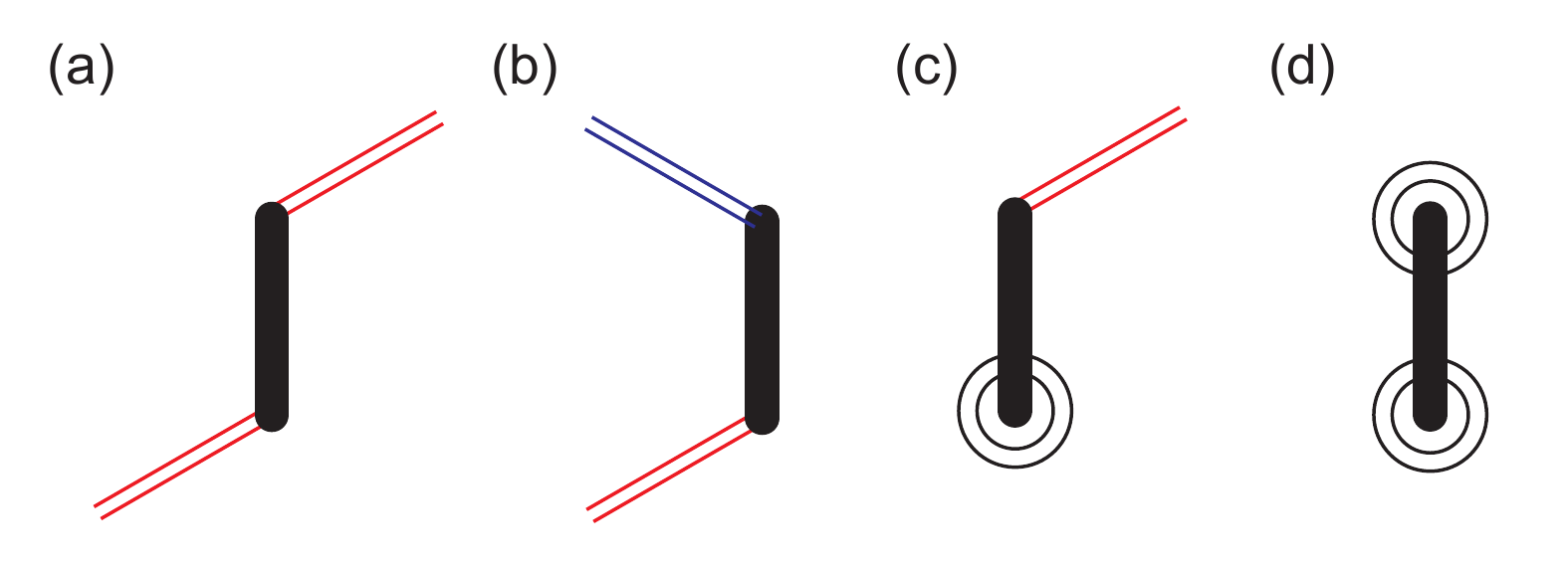}
\caption{
  Schematic pictures for the perturbation process.
  Two-fold red (blue) lines indicate that $-J_x S_i^x S_j^x$ ($-J_y S_i^y S_j^y$)
  are considered in the perturbation process twice, and double circles mean
  that $-{\bf h}\cdot{\boldsymbol{\sigma}}_i$ are considered
  in the perturbation process twice.}
\label{perturbation}
\end{figure}

Now, we focus on the $S=1$ Kitaev model
in the anisotropic-exchange and weak-field limits $|J_x|, |J_y|, |h| \ll J_z$.
In the case of $J_x=J_y=0$ and ${\bf h}=0$, two adjacent spins on the $z$-bond $\langle i,j\rangle_z$
should be fully polarized as $\ket{+1}_i\ket{+1}_j$ or $\ket{-1}_i\ket{-1}_j$,
where $|m\rangle_i$ is the eigenstate of $S_i^z$ with
the eigenvalue $m(=-1,0,1)$ at the $i$th site.
Therefore, it is convenient to introduce the pseudospin operator
$\tilde{\bf \sigma}_r$ on each $z$-bond $r$
so that $\tilde{\sigma}_r^z|\tilde{\sigma}\rangle_r=\tilde{\sigma}|\tilde{\sigma}\rangle_r$
with
\begin{eqnarray}
  \ket{\tilde{\uparrow}}_r &=& \ket{+1}_i\ket{+1}_j ,\\
  \ket{\tilde{\downarrow}}_r &=& \ket{-1}_i\ket{-1}_j,
\label{pseudo}
\end{eqnarray}
where $\tilde{\sigma}$ takes +1 ($\tilde{\uparrow}$) or -1 ($\tilde{\downarrow}$).
Now, we 
introduce $J_x$, $J_y$, and $h$ as perturbations
to derive the effective Hamiltonian for low-energy states.
Each perturbation includes $x$- or $y$-component of the spin operator,
leading to the increment or decrement of local spin quantum number $m$ by 1.
Therefore, the effective Hamiltonian appears from, at least,
fourth-order perturbation.
The perturbation processes should be represented
by the exchange couplings connected to two pseuedospins on the $x$ ($y$) bonds and
magnetic field,
which are schematically shown as the red (blue) lines and circles around a certain site
in Fig.~\ref{perturbation}.
By taking into account the corresponding contributions,
we derive the effective Hamiltonian, as
\begin{eqnarray}
  \mathcal{H}_{\rm eff} &=& - {\bf h}_{\rm eff}\cdot \sum_r \tilde{\boldsymbol{\sigma}}_r,
\label{effective}\\
  h_{{\rm eff},x}&=&\frac{7\left(J_x^2-J_y^2\right)^2}{192J_z^3}+\frac{13\left(J_x^2-J_y^2\right)h^2}{32J_z^3}\cos 2\varphi+\frac{5h^4}{4J_z^3}\cos 4\varphi,\\
  h_{{\rm eff},y}&=&\frac{13\left(J_x^2-J_y^2\right)h^2}{32J_z^3}\sin 2\varphi+\frac{5h^4}{4J_z^3}\sin 4\varphi,\\
  h_{{\rm eff},z}&=&0.
\end{eqnarray}
It is found that the effective Hamiltonian can be regarded as the free spin model
with the effective magnetic field in the $x-y$ plane.
Note that this effective field ${\bf h}_{\rm eff}$
is not directly related to the original magnetic field
since it depends on not only ${\bf h}$ but also the magnitudes of $J_x$ and $J_y$.
Then, the ground state for the effective Hamiltonian
is given by the direct product of the psueudospin state, as
\begin{eqnarray}
  |g\rangle
  &=&\bigotimes_r \frac{1}{\sqrt{2}}\left[ \frac{\left|{\bf h}_{\rm eff}\right| }{h_{{\rm eff},x}+ih_{{\rm eff},y}} |\tilde{\uparrow}\rangle_r
    +  |\tilde{\downarrow}\rangle_r\right].\label{gs}
\end{eqnarray}
In the system, the first excited energy is given by $2|{\bf h}_{\rm eff}|$
and a unique ground state is realized as far as the effective field is finite.
By contrast, when ${\bf h}_{\rm eff}=0$,
the energy scale characteristic of this effective Hamiltonian vanishes.
In the case, a higher order perturbation process should lift the macroscopic degeneracy
in the low-energy states.
Therefore, it is necessary to clarify how stable the ground state Eq.~(\ref{gs}) is
in the $S=1$ Kitaev model.
In the following, we discuss the magnetic properties in the anisotropic $S=1$ Kitaev model.

\begin{figure}[htb]
\centering
\includegraphics{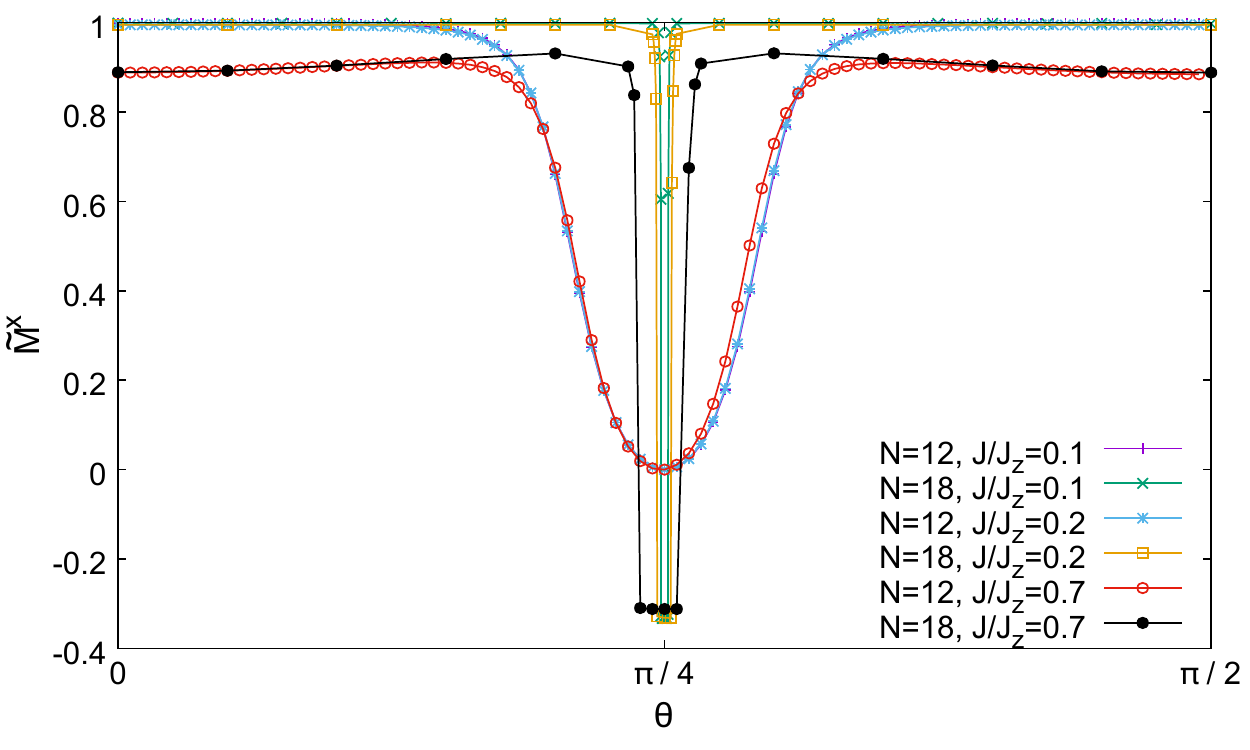}
\caption{$\theta$ dependence of the magnetization $\tilde{M}^x$
  in the ground state with $J/J_z=0.1$, $0.2$ and 0.7.
}
\label{f2}
\end{figure}

First, we consider the $S=1$ Kitaev model without the external magnetic field~\cite{Minakawa}.
When $|J_x|\neq |J_y|$ and $J_z\rightarrow\infty$, $h_{{\rm eff},x}>0$ and $h_{{\rm eff},y}$=0.
Therefore the ground state satisfies  $\tilde{\sigma}_r^x|g\rangle=|g\rangle$
for each $z$-bond.
On the other hand, different behavior may appear in the symmetric case with $|J_x|=|J_y|$
and/or away from the anisotropic limit $J_z\rightarrow\infty$.
To clarify how stable the ground state Eq.~(\ref{gs}) is
in the system with varying $J_x$ and $J_y$,
we apply the ED method to the original model Eq.~(\ref{H}) on the 12-site and 18-site clusters shown in Fig.~\ref{honeycomb}(b).
Figure~\ref{f2} shows the expectation value of magnetization
$\tilde{M}^x=\langle \tilde{\sigma}^x_r\rangle=\langle S_i^x S_j^x-S_i^y S_j^y\rangle$
for the system with fixed $J/J_z=0.1$, $0.2$, and $0.7$,
where $J_x=J\cos\theta, J_y=J\sin\theta$.
When $\theta\neq \pi/4$ ($J_x\neq J_y$), the expectation value is almost unity.
In addition, we find that $\tilde{M}^x$ takes a large value even in the case $J/J_z=0.7$,
which is far away from the anisotropic limit.
These mean that the ground state of the system is well described by the effective Hamiltonian.
We also find that, in the larger cluster, the region with $\tilde{M}^x\sim1$ becomes larger,
meaning that fourth-order Hamiltonian should capture a wide range of the parameter space
in the thermodynamic limit.
On the other hand, approaching $\theta\rightarrow \pi/4$ ($J_x\sim J_y$),
the expectation value rapidly changes from the unity.
As discussed above, the characteristic energy scale ${\bf h}_{\rm eff}$
vanishes at $\theta=\pi/4$,
and macroscopic degeneracy is not lifted by fourth order perturbation in $J_x$ and $J_y$
shown in Figs.~\ref{perturbation}(a) and (b).
In the case, other perturbation processes should be relevant.
In fact, in the 12-site (18-site) cluster,
fourth-order (sixth-order) perturbation process inherent in its periodic boundary
lifts the ground-state degeneracy, resulting in $\tilde{M}^x \lesssim 0$.
Therefore, we could not conclude the presence of such a drastic change in the magnetization $\tilde{M}^x$ around $\theta=\pi/4$ in the thermodynamic limit.


Next, we consider the effect of the external magnetic field in the system.
In general, the effective Hamiltonian Eq.~(\ref{effective}) well captures low-energy physics
for the $S=1$ Kitaev model with a large anisotropy.
On the other hand, the ground state is not trivial when the characteristic energy scale vanishes,
as discussed above.
To examine the instability of the ground state, we solve ${\bf h}_{\rm eff}=0$.
Besides the trivial solution with $|J_x|=|J_y|$ and $h=0$ discussed above,
we find the other solution as,
\begin{eqnarray}
  h_0 &=& \left[\frac{7}{240}\left(J_x^2-J_y^2\right)^2\right]^{1/4},\\
  \varphi_0 &=& \frac{1}{2}\cos^{-1}\left(-\frac{13\sqrt{3}}{4\sqrt{35}}\right)
  \sim 0.450\pi,\ 0.550\pi\ \ ({\rm mod}\ \pi).
\end{eqnarray}
Although this nontrivial solution is given
in the anisotropic limit $J_z\rightarrow\infty$,
it is unclear if this instability appears in the system
with the intermediate coupling region.
To examine how the characteristic energy scale depends
on the magnitude and angle of the applied magnetic field,
we numerically evaluate the lowest excitation gap
$\Delta$ by means of the ED method for the 12-site
and 18-site clusters.
\begin{figure}[htb]
\centering
\includegraphics[width=0.8\linewidth]{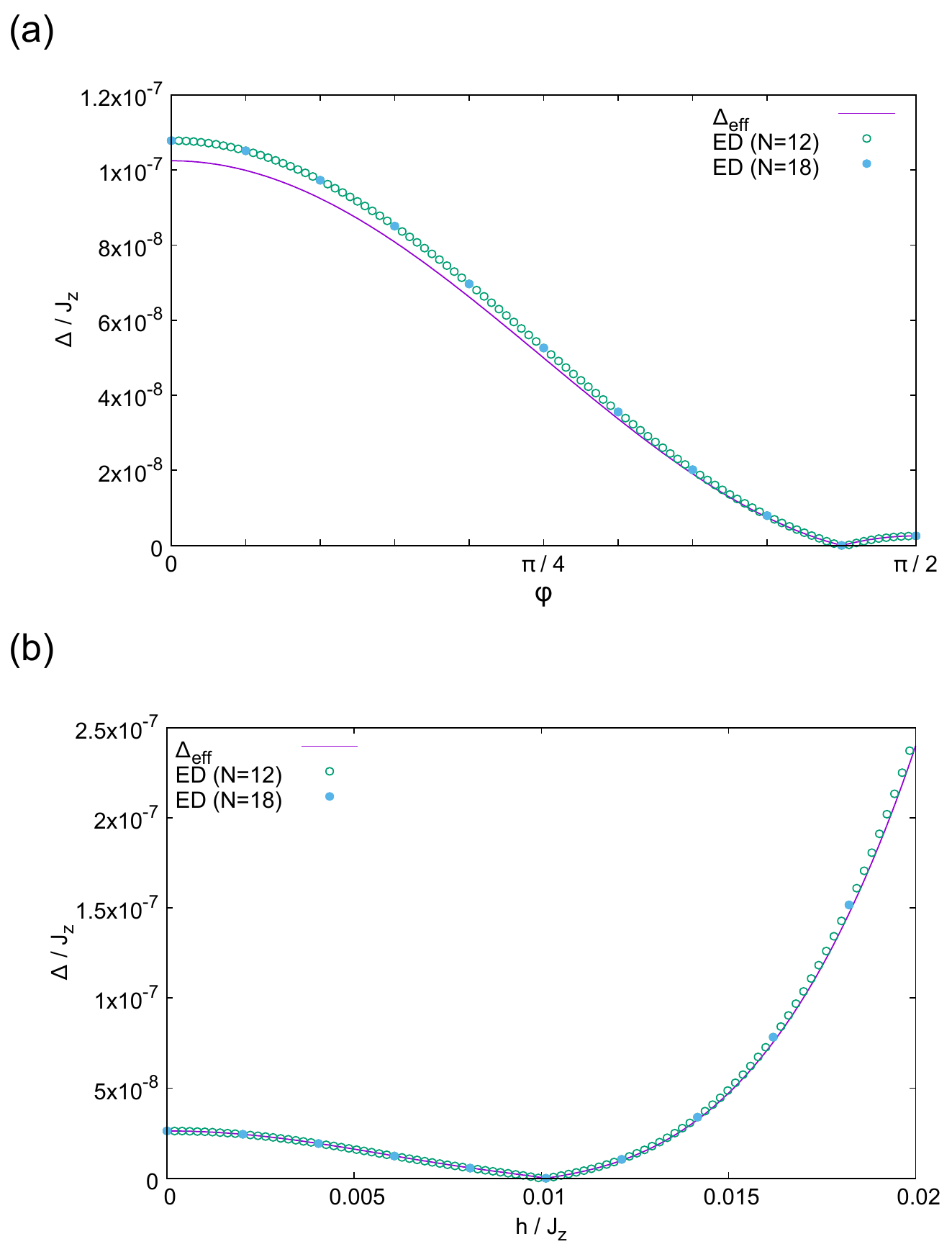}
\caption{
  Excitation gap as a function of (a) $\varphi$ [(b) $h$]
  in the system with $J_x/J_z = 0.025$, $J_y/J_z = 0.005$ and
  $h=h_0 (\varphi=\varphi_0)$.
  Open (Filled) circles represent the results obtained
  from the ED method for the 12-site (18-site) system.
  The results obtained by the fourth order perturbation
  are shown as solid lines.
  }
\label{gap}
\end{figure}
Figure~\ref{gap} shows the comparison between $\Delta$ obtained by
the ED method and that of the effective Hamiltonian, $2|{\bf h}_{\rm eff}|$.
When $J_x/J_z=0.025$ and $J_y/J_z=0.005$, we find little system-size dependence in the excitation gap.
The angle dependence of the magnetic field is shown in Fig.~\ref{gap}(a), where the magnitude of the applied magnetic field is fixed as $h=h_0$.
Away from the $x$-axis (increasing $\varphi$),
the excitation energy decreases and vanishes at $\varphi=\varphi_0$.
In Fig.~\ref{gap}(b), we show the excitation gap
in the system as a function of the magnitude of the magnetic field with a fixed $\varphi=\varphi_0$.
It is clarified that the characteristic energy vanishes
at $h=h_0$.
We also note that this energy is well reproduced by the fourth-order perturbation theory.
These indicate that, in the $S=1$ Kitaev model with the large anisotropy,
there exists a certain magnetic field ${\bf h}_0$,
where the characteristic energy scale becomes too small.
This instability is nontrivial, in contrast to the trivial solution $|J_x|=|J_y|$ and $h=0$.
Therefore, it is interesting to observe this instability
in the magnetization process experiments
in the candidate materials~\cite{Kee}.


\section{Summary}
We have studied low-energy properties of the $S=1$ Kitaev model in the anisotropic limit,
where one of the Kitaev couplings is large enough.
By performing the perturbation expansion with respect to the other Kitaev couplings and
magnetic field,
we have obtained the fourth-order effective Hamiltonian,
which is regarded as the free spin model under the effective magnetic field.
Using the ED method with small clusters,
we have discussed how stable the ground state for the effective model is
in the $S=1$ Kitaev model away from the anisotropic limit.

\section*{Acknowledgements}
This work was supported by Grant-in-Aid for Scientific Research from
JSPS, KAKENHI Grant Nos. JP19H05821, JP18K04678, JP17K05536 (A.K.),
JP16H02206, JP18H04223, 19K03742 (J.N.).

\bibliographystyle{jpsj}
\bibliography{./refs}

\begin{thebibliography}{1}

\bibitem{Kitaev}
A.~Kitaev: Ann. Phys. (N. Y.) {\bfseries 321} (2006) 2.

\bibitem{Jackeli}
G.~Jackeli and G.~Khaliullin: Phys. Rev. Lett. {\bfseries 102} (2009) 017205.

\bibitem{Baskaran}
G.~Baskaran, D.~Sen, and R.~Shankar: Phys. Rev. B {\bfseries 78} (2008) 115116.

\bibitem{S1Koga}
A.~Koga, H.~Tomishige, and J.~Nasu: J. Phys. Soc. Jpn. {\bfseries 87} (2018)
  063703.

\bibitem{S1Oitmaa}
J.~Oitmaa, A.~Koga, and R.~R.~P. Singh: Phys. Rev. B {\bfseries 98} (2018)
  214404.

\bibitem{Kee}
P.~P. Stavropoulos, D.~Pereira, and H.-Y. Kee: Phys. Rev. Lett. {\bfseries 123}
  (2019) 037203.

\bibitem{MixKitaev}
A.~Koga and J.~Nasu: Phys. Rev. B {\bfseries 100} (2019) 100404(R).

\bibitem{Minakawa}
T.~Minakawa, J.~Nasu, and A.~Koga: Phys. Rev. B {\bfseries 99} (2019) 104408.

\end{thebibliography}

\end{document}